# A Proposed Architecture for Continuous Web Monitoring Through Online Crawling of Blogs

Mehdi Naghavi[1] and Mohsen Sharifi[1]

[1] School of Computer Engineering, Iran University of Science and Technology, Narmak, Tehran, Iran
`{naghavi, msharifi}@iust.ac.ir`

## ABSTRACT

*Getting informed of what is registered in the Web space on time, can greatly help the psychologists, marketers and political analysts to familiarize, analyse, make decision and act correctly based on the society`s different needs. The great volume of information in the Web space hinders us to continuously online investigate the whole space of the Web. Focusing on the considered blogs limits our working domain and makes the online crawling in the Web space possible. In this article, an architecture is offered which continuously online crawls the related blogs, using focused crawler, and investigates and analyses the obtained data. The online fetching is done based on the latest announcements of the ping server machines. A weighted graph is formed based on targeting the important key phrases, so that a focused crawler can do the fetching of the complete texts of the related Web pages, based on the weighted graph.*

## 1. INTRODUCTION

The day to day increasing and quickly up to dated information in the Web space, provides a good opportunity for the Web users and stakeholders. One of the precious opportunities is Web monitoring .We can extract a precious treasure from the Web by Web monitoring and the information retrieval. Because of the abnormal increase in the data volume in the Web, the information retrieval has undergone a basic change. Although there is no any exact and official information about the number of the Web pages on the internet, but the number of the indexed pages announced by the Google's search engine in Feb. 2011 was more than 34 billion pages[13]. Based on the executed investigations[14], The Google in 68% of the cases, needs two days from crawling to making a page visible in search.

This research is intended to investigate the blogs monitoring and offer an architecture to implement it. To do that, the targeted crawling in the Web has been investigated and its challenges have been explained. In the continuation, the continuous monitoring of the Web sphere has been introduced as a need and the suggested architecture is offered. At the end, some related works which have been studied are presented and the conclusion is made.

## 2. MONITORING THE BLOGS

The blogs are very valuable sources whose, targeted and in time crawling would inform us about the registered events in the Web. Among the subjects done in this field, we can draw your attention to the gathering and extracting information of specific subjects like congresses[20], E-commerce[7], political viewpoints[16] the characters and the motivations of the blogger[5] discovering events in the social nets[19] tracking the changes done in the Web texts[18]. The aim of blog monitoring is discovering outstanding subjects in the Web and leading the fetching stream towards these kind of information, in a way to extract precious and important data from the Web and announcing them reasonably. The goals that could be achieved through Web monitoring are as follows:

- Finding Blogger Society Interests

- Awareness of the Various Topics Trend

- Gaining the Society Pulse

- Immediate Awareness of the Impact of Accidents and Events on Public Opinion

Then, these goals would be investigated.

The number of the internet users has been estimated to nearly two billion[9]. The best part of these users, are the readers and the bloggers. Finding the interests of the blog society can create a lot of opportunities. Among these opportunities are finding their interests in business, culture, policy and even awareness of the process of specific crimes on the internet as good examples.
In today's society, the occurrence of different events have very fast process. Specially by development of the new media technology, they would be issued in different ways and very fast .Some of these events are issued in the world of Webs. Getting informed about the process of daily events in favorite subjects could be of the other usage of the Web monitoring.

Gaining the society's pulse and interests trough the blog texts is another usage of the Web monitoring. Getting informed about the society's sensitivities and the axis around which the society's thoughts are concentrated, is of the socialists and politicians' interests. Finding the pulses of the Web user's society, can represent the pulses of the main society. Web monitoring can extract the very important axis from the Web and offers it to those in need in different fields.
If you take the Web user's society, which is a part of the human society, as the representative of a part of the public opinion, by considering their points of view, you can trace the expected society's points of view. The natural accidents and events like earthquakes, floods, tornados and tsunami, or unnatural events like wars, or social and political events like elections, putting cross country laws in action which deals with everybody in the society like increasing or decreasing salaries and taxes, would affect the society harshly. Immediate awareness about effects of these events on the society's public opinion, can give you a chance to think of a needed solution to prevent different crisis and make necessary predictions.

## 3. THE WEB TARGETED CRAWLING

Web monitoring and awareness of the registered events on the Web needs related and updated information. The targeted Web monitoring is a suitable method to achieve related information. Because of that, the focused crawlers and the blogs should be introduced and explained.

The Web crawler is a computer program, which crawls the whole universal internet with a regulated, automated method or based on a regular crawling program, and fetches their information[12]. The focused crawlers, which sometimes are called "Topical crawlers" or "vertical crawlers"[11], are of those types of crawlers which only fetch the group of information which are related to the preselected subjects.

Blogs (Weblogs) as an almost newly presented phenomenon in the World Wide Web, has made a basic change in the social relations in the Web sphere. The most important characteristics which make the blogs different from the other websites, are their universality, being interactive and their short pace getting up to dated. Up to the year 2009, more than 133 million blogs by the search engine Technorati, special search engine fore blogs, have been indexed [3], and more than 900000 new blog posts have been registered per day [6].

## 4. THE CHALLENGES OF THE BLOGS' TARGETED CRAWLING

To monitor the Web, it is necessary to targeted crawl the Web space. The targeted crawling of the blogs faces following challenges, which will be mentioned later.

- Short Pace Updating of the Blogs
- The Blog's Transitory Nature
- The Problem of Classifying Blogs
- The Spam Blog and Linking to it
- Using the Informal Language in Blogs
- The Existence of Millions of the Empty Blogs without Posts
- The Detachment of the Documents in the Blogs

Considering the too many blog writers, and their interests in putting the news and new information in the blogs, which are called 'posts', the up to dating pace for active blogs would be very short. So, as soon as any change in the blogs occur, and any new post is made, we should retrieve their new information in order to make use of.

The information on the blogs are time sensitive, and their importance decreases or would be ineffective through time. In most cases, retrieving the blog post's information and processing it in a short time is very important.

Classifying blogs by human being or even by machines is very difficult. It is difficult for man, because there is a great amount of blogs in front of him and he needs a lot of human force to classify them based on human investigation, wisdom and knowledge. It is difficult for machines, because they do not follow a certain structure and are based on people's tastes and interests and in many cases they are written irregularly, as wanderers and unwisely.

One of the difficulties in crawling blogs are the spam pages. Based on the Technorati report, the number of the new blog spam in each day reaches to 11000 pages[20]. The spam links are made by the explanation and the answers which are supported actively. The spam made by explanations are easily created in the blogs and links are attached to them. These links create false relations,

International Journal of UbiComp (IJU), Vol.3, No.1, January 2012which carry no meaningful load. These kinds of relations deceive the crawlers, specially the focused crawlers.

Using the informal or spoken form and conversational language in blogs cause a great variety in words and their meanings. The blog language contains numerous vocabulary from the technical field of computer and also words with uncommon hand writings in the Web .These kind of writings contain relatively lots of typing and writing mistakes. These cases make the understanding of the word's meanings and targeted monitoring on the Web very difficult.
With the development of blogs, a great deal of blogs have been produced by active users in this field, which have been left useless. So that the crawlers should investigate and fetch millions of empty pages and through them away. This causes to destroy a great deal of sources like bandwidth and processors.

The pages in the blog's normal environment are connected to each other through sub pages, and that way, they keep their meaningful relations to each other. In social nets including blogs, the documents are often detached[8] and they may not have a defined relation with the related hierarchy. This problem will come up in a targeted crawling on a certain subject.

## 5. CONTINUOUS MONITORING OF THE BLOG SPACE

Monitoring the blog in order to get aware of the events, registered in them, is a method for getting informed quickly on important events in the society. If we focus only on the blog instead of the whole Web, the working domain of the crawler from a 34 billion page space will be focused on a 133 million page space[6]. Also, crawling this great volume of blogs for special applications is impossible. Thus, for more focusing, the target space should be more limited and up to dated blogs should be targeted. In the continuation, a definition for continuous monitoring of blog space will be given.

**Definition**: Blog monitoring means continuous and online investigation of the blogs and extracting and offering the important events and occurrences written on them, in acceptable time.
The purpose of the continuous and online investigating blogs is to study the blogs' contents and getting aware of the news and new events registered in them. To distinguish the news and important events, we can make use of distinguishing very important key phrases which have specific characteristics. Very important key phrases are those phrase, by which we can distinguish the importance of the news and the value of the events registered in them.

A key phrase is a phrase that at least contains two or more words [4]. The key phrases do not necessarily follow the common structures of a language, like having verbs, subjects and subject predicates. In order to observe more consistency between the similar key phrases, we should possibly divide them into their main and important particles. Stop words like prepositions, auxiliary verbs, different pronouns and some adjectives could be deleted  from the phrase  and then do the processing on the cut downed  phrase.

To monitor a Web we should distinguish the very important key phrases and extract them. We can introduce many different criteria for a key phrase to be very important or less important. Some of criteria listed below and will be explained.

- The Number of Repetition of a Key Phrase

- Being Active and Updated of a Blog Containing the Key Phrase

14

- The Characteristics of the Blog Writer
- Input and Output Links of a Blog

The repetition of a key phrase can be used as one of the acceptable criteria for showing its importance. The more the number of repetition of a phrase, the more important the phrase is.

The other indicator which could be used for determining the amount of the importance of key phrases, is the key phrase containing blog's being active and up to dated .To determine this indicator, we can use the number of blog visitors and the notes left to answer blog's post or the time paces of updating the posts.

The blogger's characteristics could be an indicator for showing the amount of the importance of a key phrase. So based on this the blog writers could be categorized, based on their activity and influence as different groups like "Active-Influential", "Inactive-Influential", "Active-Non Influential" and "Inactive -Influential" [1].

The number of links entering a page and exiting from it can be another criterion for the phrases of that page. There is a strong relation between the number of input and output links with the amount of importance of a page and the key phrases in it. The normal pages and the pages which have many visitors can be distinguished by this indicator[22].

If we consider the criterion of the number of repetition as the main indicator, regardless the criterion of hierarchy, links referring can give more weight and value to the criterion of repetition. The indicator of up to dating is highly imbedded in the criterion of number of repetition. The high number of repetition of a key phrase is the sign of the related blog's being up to dated and active, because the importance of a key phrase makes more users to make comments about it and leave notes on the posts. Thus, in order to speed up the key phrase distinguishing algorithm, the repetition indicator has been used, and the number of its words has been taken 2 or 3.

## 6. BLOGS MONITORING LIMITATIONS

In offering the blog monitoring architecture, there are limitations like the access time to the results, the processing capacity and the bandwidth, knowing about which will help us to develop a suitable architecture. These limitations will be explained later and then considering them, the online blog monitoring architecture for continuous monitoring of the Web space will be introduced.

One of the characteristics of the target space monitoring, will be doing it in a certain, acceptable and limited time. If the results get ready after the expected time, the obtained data loses its application and value and will be useless. Also, there should not be any pause in doing the monitoring, because of different reasons, like investigating the previous results or great volume of information.

The process of Web monitoring is not a general process and has specific and limited customers. Web monitoring could be done with different goals, like social, economical, cultural, political or other goals. The pursuers of these limited goals are those organizations that have a certain, known and limited capacity and for having access to these kind of systems, they face serious limitations. It is not possible for these kinds of customers to add up easily to their software and hardware to extol their processing capacity. The offered architecture should be able to consider these limitations and should be designed in a way that could be implemented with these kinds of processing capacity limitations of the customers.

The bandwidth capacity is also a case that, because of imposing high expense, is a kind of limitations for customers. The common crawlers use a great amount of the bandwidth to fetch the data. The mentioned architecture should be offered in a way that could be implemented either with limited connecting bandwidth or with unlimited bandwidth. In other words, it should be able and dynamic to work suitably with changing bandwidth.

## 7. THE ARCHITECTURE OF THE BLOG ONLINE MONITORING

Figure 1 shows the architecture of the focused crawling for discovering the professional events based on the user's interests[17].

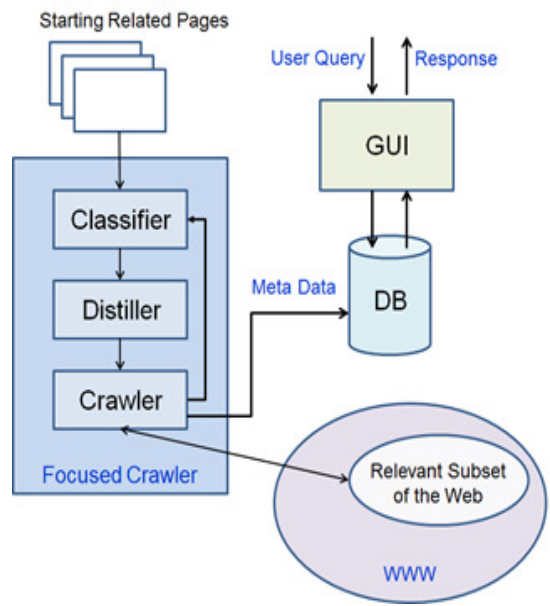

Figure 1. Architecture of the focused crawler for professional events

In this research, The Vector Space Model or VSM Algorithm[2] has been used and some suitable algorithms have been investigated for being categorized in focused crawlers, like SVM[10] and Bayesian Rule [15]. When the page is categorized as the related topic, the existing addresses in this page are used for the continuation of the crawling and otherwise crawling of this page will be stopped.

The suggested architecture of this article has been offered by investigating the above architecture and also architectures and methods which have been mentioned in the related work section. In the suggested architecture, instead of using the prepared initial seed, the online producer of the addresses' list is used. The most important characteristics of Web monitoring is that the Web sphere crawling is online. The suggested architecture should be perfectly able to support that, so that the continuous Web sphere monitoring could be done.

Figure 2 shows the suggested blog's online crawling architecture for monitoring the web sphere. In order to cover the existing limitations and mentioned goals, the suggested architecture has been offered in three layers. The first layer of the architecture, is responsible fore providing the initial

16

seeds for the crawler. Unlike the common crawlers which start with one or more Web pages like Dmoz website, and then the extracted links from them will determine the crawling route, in this suggested architecture, the initial seeds are continuously produced and given to the crawler online and based on the latest events. To retrieve the lately changed blogs, the ping mechanism[8] is used. In this technology, the active websites will, immediately, inform the ping servers about the latest changes in their websites. These servers also will, continuously, issue these changes online. After receiving the list of the changed websites, it compares it with the blogs addresses list that already prepared, and in case it is among them, it gives it to the second layer as initial seeds.

The second layer which is the crawler of the blog's summary, based on the initial seeds of the first layer, takes an address and using the RSS (Rich Site Summary) technology starts fetching the related blog summary. Then, based on an analysis with the semi vsm algorithm, recognizes and extracts very important phrases which contain at least 2 and at most 3 words, and based on them and using the existing links in that page, will form a weighted graph. This graph's nodes will show the addresses of the Web and this graph's edges `show the estimated weight of the key phrases in the destination nodes. Since this graph has been formed based on the weight of the amount of the key phrase repetition, it will be directed and targeted to the very important key phrases. In this layer, the received data is not accumulated, but it is processed simultaneously. Immediately after receiving the summary related to an address, the processes related to its analysis is done and after it is finished, we go to the next address. The crawler of this layer, unlike the crawler's common trend, does not use the new addresses, he himself has extracted, and does not enter them in the fetching process. This will keep the online characteristics of the crawling. Crawling a summary of a part of a blog is not enough at all for crawling the Web. The gathered data lacks many things which causes the formed graph lack enough precision. To compensate this defect, the third layer has been developed.

In the third layer, a focused crawler, based on the existing graph, tries to fetch the related pages. This fetching includes the complete texts of the pages. In order to make this step's function more effective, fetching some of the meta texts, like photos, voices, and movies is not done. The fetcher in this layer which can be separately implemented on some parallel machines, would fetch the pages based on the existing graph's information and will put it in related reservoir. After fetching the pages, a text analyser will start to correct the graph, based on related regulations, word glossary, spam page filters, and blog distinguishers, so that the new page fetching target domain would be determined with more concentration.

The three layers of the offered architecture would complete each other and their cooperation, will make the crawling online, speed up, and focus on very important related key phrases. At the end of the crawling process, in the third layer, the most related pages with very important key phrases will be gathered, and made ready for the next process.

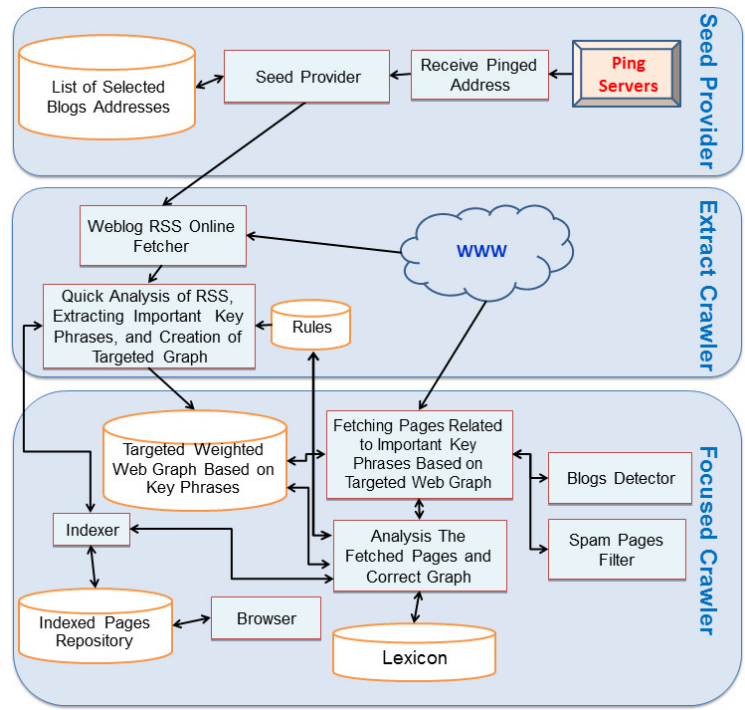

Figure 1. Architecture of the blog online crawling for monitoring the Web

## 7. RELATED WORK

Huang and his colleagues[7] have offered a semantic approach for implementing a focused intelligent crawler by hyperlink analysis and domain ontology with focusing on E-commerce. Agarwal and his colleagues [1], have offered a system to help the sociologists to track, analyze, word monitor in the blogs and categorize their writers to "Active-Influential", "Inactive-Influential", "Active-Non Influential" and "Inactive -Influential". Ye and his colleagues[21] have investigated the problems and its challenges in online graph crawling of social networks. Distinguishing and discovering events in social nets using word graphs, is another subject investigated by Sayyadi and his colleagues[19], in Microsoft Co. Oh and his colleague[16] have been involved in investigating a ranking system based on a learning model of the blog post's political viewpoints. Pathak and his colleague[18], have presented an intelligent method for monitoring the changes done in the websites.

## 7. CONCLUSION

In the process of Web monitoring, the most important goal is fetching the related data in a certain time. In this article, A three layer architecture is introduced, which, fetches new and important information in the blogs in a certain time pace, using two summary and focused crawlers. The first type of the crawler, using the addresses list of the newly changed blogs, which has been prepared by Ping Technology, tries to fetch data of the page summary, by the RSS technology, and then by analyzing the fetched data quickly, starts to extract the key phrases and based on that, will form a weighted graph. The second type crawler, which is a focused crawler, starts to fetch the complete pages, based on the weighted graph, and using a text analyzer based on the new information starts to correct the graph.

We can consider different goals for Web monitoring on fetched pages, which could form our future new research tittles. Cases like distinguishing the issuing source of the news, recognizing the news publication network, distinguishing correct and false news, and gossipy, can be our future research projects.

## Authors

Mehdi Naghavi is a PhD candidate at Iran University of Science and Technology. His doctoral research is directed toward identifying and addressing challenges in continuous web monitoring. He received his M.Sc. in Computer Engineering (Computer Architecture) from Amirkabir University of Technology, Tehran, Iran, in 1995 and received his bachelor's degree in Electrical Engineering (Hardware) from Sharif University of Technology, Iran, Tehran, in 1991. His areas of interests include system software, distributed systems, Information retrieval, and online crawling.

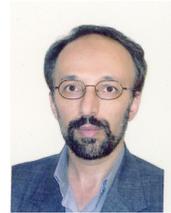

Mohsen Sharifi is an Associate Professor of Software Engineering at the Computer Engineering Department of Iran University of Science and Technology. His main interest is in the development of distributed systems, solutions, and applications, particularly for use in various fields of science. He has developed a scalable high performance and high available computing (HPAC) cluster solution running on any number of ordinary PCs for use in scientific domains requiring high performance and availability. The development of a true distributed operating system is on top of his wish list. He received his B.Sc., M.Sc. and Ph.D. in Computer Science from the Victoria University of Manchester in the United Kingdom in 1982, 1986, and 1990, respectively.

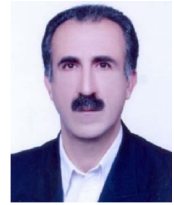